\documentclass[aps,prl,twocolumn,showpacs,superscriptaddress]{revtex4}

\usepackage {graphicx}
\usepackage{epsf}

\begin{document}

\title{High resolution probe of coherence in low-energy charge exchange collisions with oriented targets}

\author{A. Leredde}
\affiliation{LPC Caen, ENSICAEN, Université de Caen, CNRS/IN2P3, Caen, France}
\author{X. Fl\'echard}
\affiliation{LPC Caen, ENSICAEN, Université de Caen, CNRS/IN2P3, Caen, France}
\author{A. Cassimi}
\affiliation{CIMAP, CEA - CNRS - ENSICAEN, BP 5133, F-14070, Caen cedex 5, France}
\author{D. Hennecart}
\affiliation{CIMAP, CEA - CNRS - ENSICAEN, BP 5133, F-14070, Caen cedex 5, France}
\author{B. Pons}
\affiliation{CELIA, Univ. Bordeaux - CNRS - CEA, F-33405 Talence, France.}

\date{\today}

\begin{abstract}
The trapping lasers of a magneto-optical trap (MOT) are used to bring Rb atoms into well defined oriented states. Coupled to
recoil-ion momentum spectroscopy (RIMS), this yields a unique MOTRIMS setup which is able to probe scattering dynamics, including 
their coherence features, with unprecedented resolution. This technique is applied to the low-energy charge exchange processes
Na$^+$+Rb($5p_{\pm 1}$) $\rightarrow$ Na($3p,4s$)+Rb$^+$. The measurements reveal detailed features of the collisional interaction
which are employed to improve the theoretical description. All of this enables to gauge the reliability of intuitive pictures
predicting the most likely capture transitions.
\end{abstract}


\maketitle

Apprehending the dynamics of nonadiabatic processes in atomic and molecular physics is not an easy task. The initial
and final states have generally complex fine structures and even in the case of well-defined boundary conditions, multiple 
pathways can drive the dynamics from the beginning to the end of the process. Beyond the identification of the reaction 
routes, their relative coherence must explicitely be considered since it monitors the eventual occurence of interferences 
and determines the final outcome of the reaction. The degree of coherence, that our (basically classical) intuition cannot easily grasp, 
is generally revealed by quantum mechanical or semiclassical calculations. However continuous advances in coincidence and probe techniques
have opened the way to 'quantum mechanical complete' coherence experiments \cite{Andersen88} which are able to measure not only the reaction 
probabilities but also the relative phase of the transition amplitudes. Such experiments have been performed for low-energy
ion-atom collisions \cite{Thomsen96,Roller-Lutz93,Salgado97,Thomsen99,Roller-Lutz00,Dowek02}, electron impact excitation of helium 
\cite{Mikosza,Cvejanovic} and photon-induced ionization of krypton atoms \cite{Goulielmakis10}. Nevertheless, in most of the cases 
involving heavy particle impact, the experimental resolution did not allow to scrutinize the details of the scattering patterns 
predicted theoretically. 

We are interested in ion-atom collisions in the keV impact energy range where the most probable inelastic process
is electron transfer from the target to the projectile \cite{Bransden}. We have recently demonstrated that the transfer dynamics 
in Na$^+$+Rb($5s,5p$) collisions can be probed with unsurpassed accuracy by coupling cold-target recoil-ion spectroscopy 
(COLTRIMS \cite{Dorner00}) to laser-cooled atomic targets trapped in a magneto-optical trap (MOT) \cite{Vanderpoel01,Turkstra01,Flechard01}. 
Here we show that the MOT trapping laser pulses can further be employed to bring the target into an oriented initial state, yielding a novel
MOTRIMS setup which enables to picture the charge exchange dynamics, including its coherence features, with unprecedented resolution. 
We still consider Na$^+$+Rb collisions, but optical pumping from a circularly polarized laser pulse selects the entry state as $5p_{\pm 1}$. 
The high-resolution angular differential cross sections (DCS) are compared to molecular orbital close-coupling (MOCC) calculations performed 
in the framework of the single active electron (SAE) approximation \cite{Leredde12}. This comparison reveals subtle features of the 
projectile-target interaction which monitors the state-selective scattering, and further allows us to mark out the accuracy of the widely used 
SAE approach. Additionally the DCS's show a pronounced and finely resolved asymmetry between left- and right-handed scatterings, related to 
the coherence of the charge-exchange process from the selected oriented state. These orientation effects are theoretically studied from both 
classical and quantum perspectives.  

\begin{figure}
\includegraphics[width=\columnwidth]{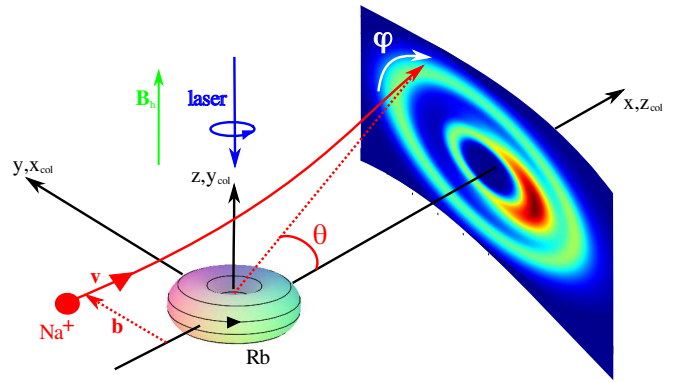}
\caption{A homogeneous magnetic field $B_h$ sets the quantification $z$-axis and optical pumping leads to magnetic sublevels of the target state with well defined 
hyperfine $m_F$ quantum numbers, depending on the handedness of the laser pulse. 
We illustrate this for the Rb($5^2 P_{3/2},F=3,m_F=+3$) $\equiv$ Rb($5p_{+1}$) case. Na$^+$ impinges on the oriented target impact with initial velocity $\bf{v}$
and impact parameter $\bf{b}$.}
\label{orientation}
\end{figure}

%Experimental setup
Our MOTRIMS setup has been described in details elsewhere \cite{Leredde12,Blieck2008}. We therefore focus hereinafter on the transformations we made to 
merge efficient orientation and RIMS techniques within a MOT. 

The high resolution performances of our MOTRIMS setup rely on a transverse extraction of the recoil ions coupled to a fast switch-off of the trapping magnetic 
field during data counting \cite{Leredde12}. The field-free requirement is hardly compatible with a definite orientation of the target states since 
weak parasite magnetic fields, including the terrestrial one, can randomly influence the state alignment which has to be first strictly fixed. The total 
kinetic momentum of the target is thus unambiguously aligned, for both the ground Rb($5^2 S_{1/2},F=2$) and excited Rb($5^2 P_{3/2},F'=3$) states, by using a 
homogeneous magnetic field $B_h$ along the $z$-axis which is taken as the quantization direction (see Fig. \ref{orientation}). $B_h$ is provided by Helmotz coils, 
and its magnitude is 2 Gauss, which is large enough to depreciate any perturbing magnetic field. To compensate the action of $B_h$ during the trapping period, 
one of the MOT anti-Helmotz coil is shifted up by 4 mm so that the field-free requirement is obeyed at the center of the collision chamber. 

To bring the atoms Rb($5^2 S_{1/2},F=2$) and Rb($5^2 P_{3/2},F'=3$) into oriented $m_F=+2$ and $m_F'=+3$ magnetic sublevels, respectively, we optically pump them 
using a left-handed circularly polarized laser beam, coming along the $-z$ direction, that induces $\sigma^+$ transitions. The opposite orientation is obtained using a 
right-handed laser beam. 
The polarizing laser beam (PLB), coming from one of the trapping lasers, is diffracted by an AOM (Acousto-Optic Modulator) and tuned slightly to the red of
the $\left( 5^2 S_{1/2},F=2 \right) \rightarrow \left( 5^2 P_{3/2},F'=3 \right)$ transition. This AOM can be switched \textit{on} and \textit{off} to 
control the status of the PLB.
A half-wave plate coupled to a PBSC (Polarizing Beam Splitting Cube) provides a linearly polarized beam with tunable intensity. The beam is circularly polarized 
using a quarter-wave plate and directed towards the target. 
A counter propagating laser beam, obtained by a retro-reflection at the exit side of the collision chamber, reduces net momentum transfer from the PLB to the target atoms. 
In spite of this retro-reflected beam, the cold cloud is still warmed up and lost after a few ms with a time constant depending on the intensities and alignment of the 
laser beams.
An incoming laser beam with a power of 1 mW/cm$^2$ was thus used to provide a target with a large excited fraction 
without pushing the cold atomic cloud outside of the collision region. 

\begin{figure}
\includegraphics[width=\columnwidth,clip]{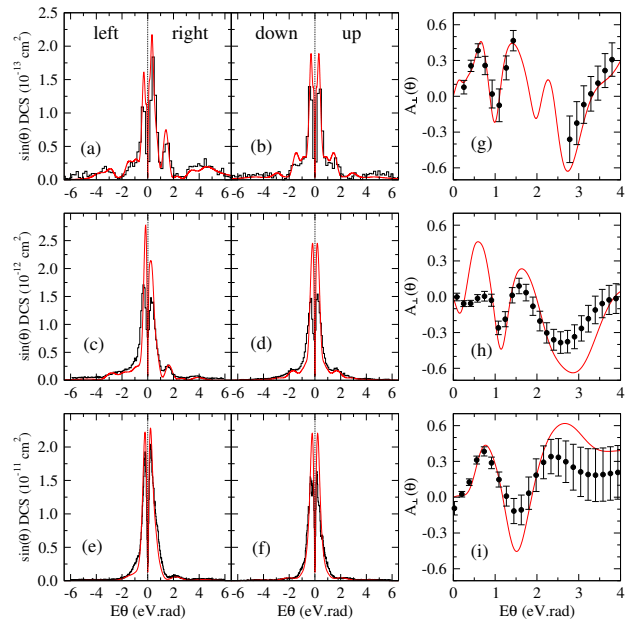}
\caption{Weighted DCS's relative to the charge-exchange reaction Na$^+$+Rb($5p_{+1}$) $\rightarrow$ Na($3p$)+Rb$^+$ at $E=1$ (a-b), 2 (c-d) and 5 (e-f) keV, as functions of
$E\theta$. The histograms refer to the measurements while the continuous (red) lines correspond to MOCC calculations. 
In (g-i) are displayed the left-right coherence parameters $A_\perp(\theta)$ defined in (\ref{Aperp2}) for $E=1$, 2 and 5 keV respectively.}
\label{fig2}
\end{figure}

The polarization speed and orientation yield have been characterized by means of absorption and fluorescence techniques, respectively. We found that
more than 95$\%$ of the atoms are conveniently oriented within a time interval shorter than 5 $\mu$s. 
Experiments were performed with oriented Rb($5^2P_{3/2},F=3,m_F=\pm3$) $\equiv$ Rb($5p_{\pm 1}$) targets at 5, 2 and 1 keV. Acquisition times for both right- and left-handed 
orientations of the target were of about 24 hours for each energy. 
We present in Fig. \ref{fig2}(a-f) the DCS associated to the main Na$^+$+Rb($5p_{+1}$) $\rightarrow$ Na($3p$)+Rb$^+$ charge-exchange reaction. The angular resolution
achieved ranges from $\Delta \theta$=70 $\mu$rad at 1 keV to 28 $\mu$rad at 5keV. 
The measurements are compared to MOCC calculations detailed in \cite{Leredde12}. Briefly MOCC is a semiclassical approach which combines a classical
description of nuclear motion and a quantum-mechanical description of electron transitions. Only nonadiabatic transitions involving the Rb valence electron are
considered and both the target and projectile atomic cores are assumed to remain frozen throughout the collision. Parametric model potentials then
represent the interactions of the active electron with the ionic cores \cite{Leredde12}, and the description of the core-core interaction is first 
restricted to its coulombic repulsive part, as usual in treatments of collisions with dressed target and projectile (see e.g. 
\cite{Dubois93,Machholm94}). MOCC employs real orbitals $\left\{p_x,p_y,p_z\right\}$ and computes the scattering amplitudes $T^{col}_{p_{x,y,z}\rightarrow p_{x,y,z}}(\theta)$ 
in the collisional frame ($x_{col},y_{col},z_{col}$) where the $z_{col}$- and $x_{col}$-axis are defined by the projectile beam and the impact parameter directions respectively. 
The amplitudes $T^{col}$ are then rotated to correspond to initial and final oriented states 
defined in the photonic reference frame ($x,y,z$), as documented in e.g. \cite{Machholm94,Sidky02} and illustrated in Fig. \ref{orientation}. This yields
\begin{eqnarray}
T_{p_{+1} \rightarrow p_{\pm 1}}(\theta,\varphi)= \mp \frac{1}{2} T^{col}_{p_z\rightarrow p_z} - \frac{1}{2} \cos^2(\varphi) T^{col}_{p_x\rightarrow p_x} \nonumber \\
+ \frac{1}{2} \sin^2(\varphi) T^{col}_{p_y\rightarrow p_y} + \frac{i}{2}\cos(\varphi)\left[ T^{col}_{p_z\rightarrow p_x} \mp T^{col}_{p_x\rightarrow p_z}\right] \nonumber \\
T_{p_{+1} \rightarrow p_{0}}(\theta,\varphi)= \frac{\sqrt{2}}{2} \sin(\varphi) T^{col}_{p_z\rightarrow p_x} + \frac{i\sqrt{2}}{4} \sin(2\varphi) \nonumber \\
\left[ T^{col}_{p_x\rightarrow p_x} + T^{col}_{p_y\rightarrow p_y}\right] \label{T} 
\end{eqnarray}
We do not resolve experimentally the $m$-contributions to the $3p$ DCS and therefore display in Fig. \ref{fig2}(a-f) the weighted DCS 
$\sin(\theta)\sigma_{p_{+1} \rightarrow 3p}=\sin(\theta)\left(|T_{p_{+1} \rightarrow p_{+1}}|^2+|T_{p_{+1} \rightarrow p_{0}}|^2+|T_{p_{+1} \rightarrow p_{-1}}|^2\right)$. 
Furthermore, for sake of quantitative comparison between experiments 
and theory, the DCS is presented in terms of its four main contributions to which we refer to as left ($\varphi=0$), up ($\varphi=\pi/2$) , right ($\varphi=\pi$) 
and down ($\varphi=3\pi/2$) \cite{note}. The calculations are convoluted according to the experimental resolution $\Delta \theta$ previously detailed. 

The agreement of experimental and computed DCS's is very good, and this is even more satisfactory as we normalize only the whole measured signal to 
the total computed cross section. Because of the reflexion symmetry with respect to the ($x,y$)-plane, the up and down contributions are identical. This is not the 
case for the left and right contributions: the rotation of the electron flow inherent in the initial oriented state breaks the symmetry of left ($y>0$) and 
right ($y<0$) scatterings (see Fig. \ref{orientation}). The left/(right+left) total ratios, $R$, derived from the 
measurements and MOCC calculations are in striking agreement; for instance, $R^{MOTRIMS}=37.7\%$ and $R^{MOCC}=37.5\%$ at $E=1$ keV. Furthermore,
and importantly, the exceptional angular resolution reached by our improved setup ($\Delta\theta \sim 70$ $\mu$rad at 1 keV) enables to resolve the oscillatory 
patterns of the DCS and associated ($\theta$-dependent) left-right asymmetry. This would have been clearly impossible with conventional setups which rather deal with 
$\Delta\theta \sim$ 1 mrad. The $\theta$-dependent left-right asymmetry, defined as $A_\perp(\theta)=\frac{\sigma_{p_{+1} \rightarrow 3p}
(\theta,0)-\sigma_{p_{+1} \rightarrow 3p}(\theta,\pi)}{\sigma_{p_{+1} \rightarrow 3p}(\theta,0)+\sigma_{p_{+1} \rightarrow 3p}(\theta,\pi)}$, 
is nothing else but a direct measure of the importance of interference effects between 'radial' ($p_z \rightarrow p_z$,$p_x \rightarrow p_x$) and 
'rotational' ($p_z \rightarrow p_x$,$p_x \rightarrow p_z$) collisional transition pathways since one obtains from equations (\ref{T})
\begin{equation} 
A_\perp(\theta)=\frac{2 \ \text{Im}{(T_{p_z \rightarrow p_z}T^*_{p_x \rightarrow p_z}-T_{p_x \rightarrow p_x}T^*_{p_z \rightarrow p_z}})}
{|T_{p_z \rightarrow p_z}|^2+|T_{p_x \rightarrow p_x}|^2+|T_{p_z \rightarrow p_x}|^2+|T_{p_x \rightarrow p_z}|^2}
\label{Aperp2}
\end{equation}
In other words, $A_\perp(\theta)$ gives direct access to the degree of coherence of the Rb($5p_{+1}$) $\rightarrow$ Na($3p$) charge exchange mechanism. 
$A_\perp(\theta)$ is presented in Fig. \ref{fig2}(g-i) for $E=1,2$ and 5 keV. At $E=1$ keV, $A_\perp(\theta)$ could not be safely derived from the measurements 
between $E\theta=1.5$ and 2.5 eV.rad because of excessively noisy left and right signals.  
The overall agreement of the measured and predicted coherence pictures is satisfactory. As $E$ decreases, the interferential pattern presents more
structures within a fixed $E\theta$ range which, according to the classical description of scattering, should mark the range of involved impact parameters 
$b$ through $b=1/E\theta$ \cite{Bransden}. This behaviour stems from the fact that the phases of the collisional transition amplitudes are inversely proportional to the impact
velocity $v$ \cite{Leredde12}, which enhances the number of constructive and destructive occurences in the $T$-cross products of (\ref{Aperp2}) within
a fixed $b$-range as $E$ decreases. 

\begin{figure}
\includegraphics[width=\columnwidth,clip]{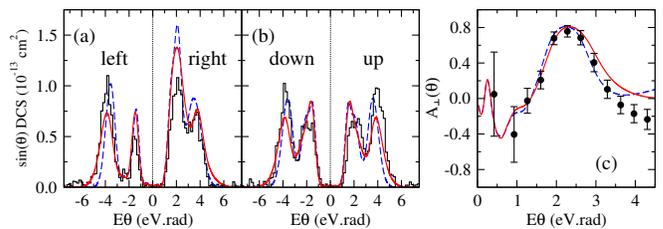}
\caption {Same as Fig. \ref{fig2} for Na$^+$+Rb($5p_{+1}$) $\rightarrow$ Na($4s$)+Rb$^+$ at $E=1$ keV. The dashed (blue) and continuous (red) lines correspond to MOCC calculations accounting or not for high-order multipole core-core interactions.}
\label{fig3}
\end{figure}

In spite of the rather low statistics inherent in ($\theta,\varphi$)-differential measurements, our setup has allowed us to reliably extract from
the raw data the signal associated to the secondary capture channel Na$^+$+Rb($5p_{+1}$) $\rightarrow$ Na($4s$)+Rb$^+$ at $E=1$ keV. To our knowledge 
no previous measurements were able to scrutinize secondary transfer dynamics at the coherence level. The DCS and corresponding 
$A_\perp(\theta)$ are presented in Fig. \ref{fig3}. MOCC and MOTRIMS lead to a double ring structure of the DCS which is mainly imprinted by 
$5p_0 \rightarrow 4s$ transitions occuring at rather small internuclear distances $R < 12$ u.a. But while the measurements yield a maximum outer ring, 
MOCC leads to the opposite. We demonstrate in Fig. \ref{fig3} that this is due to the inadequate representation of the core-core interaction 
in terms of purely net coulombic repulsion $1/R$.  We have indeed implemented improved MOCC calculations
in which the mutual polarization of the cores is accounted by the attractive ion-dipole $-(\alpha_d^{Rb^+}+\alpha_d^{Na^+})/2R^4$ and dipole-dipole 
$-(\alpha_q^{Rb^+}+\alpha_q^{Na^+})/2R^6$ potentials, where $\alpha_d$ and $\alpha_q$ are the dipole and quadrupole polarizabilities 
of the ionic cores \cite{polarizabilities}. These additional terms are naturally cancelled at small $R$, where the total core-core interaction
must be repulsive, since we introduce in our MOCC calculations a cut-off function which prevents the system from entering the $R$-domain
where the cores overlap \cite{Leredde12}. It is clear in Fig. \ref{fig3}(a,b) that accounting for high multipole terms in the core-core interaction
significantly affects the $4s$ scattering and improves the agreement of calculations and measurements. In the case of capture into the $3p$ shell, which mostly occurs
at large $R$ ($R \gtrsim 15$ a.u.), changes with respect to the net coulombic behaviour were found to be small.

The asymmetry parameter associated to charge exchange into $4s$ directly measures the inteferences between radial $p_z \rightarrow s$ and
rotational $p_x \rightarrow s$ pathways, since $A_\perp(\theta)=2 \ \text{Im}{(T_{p_z \rightarrow s}T^*_{p_x \rightarrow s})}$. 
The improved MOCC calculations enhance the agreement of the computed $A_\perp$ with its experimental counterpart between $E\theta=2$ and 3 eV.rad. 
Beyond this range, both results deviate. In fact the dipolar and quadrupolar core-core potentials introduced in the improved calculations 
induce strong oscillating phases in the expression of the scattering amplitudes $T$. These phases, which respectively behave as $1/(vb^3)$ and $1/(vb^5)$, 
are essential to correctly reproduce the relative height of the DCS maxima as previously observed, but erode the outer part of the DCS upon integration on $b$. 
This results in a slight mismatch of the positions of the outer DCS maxima and leads to the deviation of $A_\perp$ at large $\theta$'s. 
Improving further the theoretical calculations is beyond our present capabilities for such a complex system as Na$^+$+Rb: prohibitive ab initio calculations should be
performed to explicitely represent the molecularization of the ionic cores at small $R$, well beyond the present model potential with
frozen atomic cores SAE description. This notwithstanding, the present improved MOCC calculations show a whole satisfactory
agreement with experiments. 

\begin{figure}
\includegraphics[width=\columnwidth]{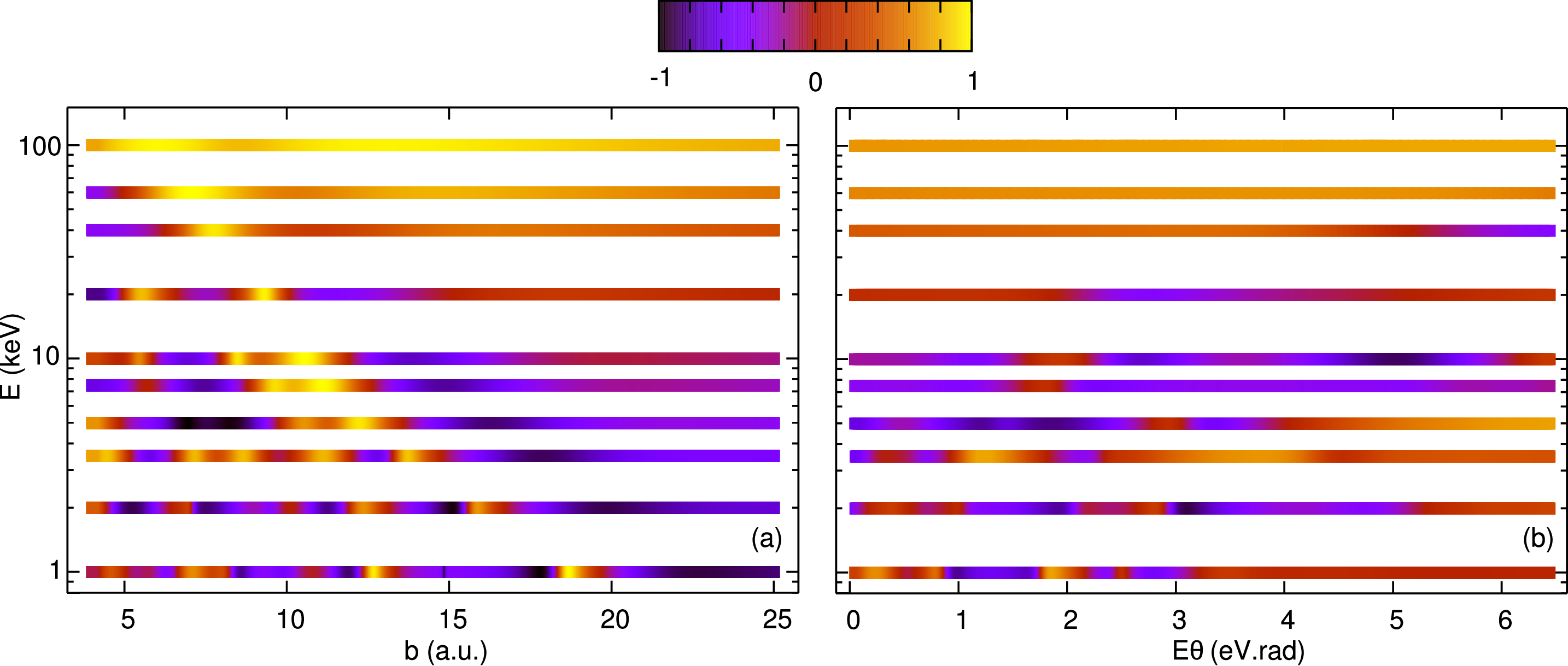}
\caption{(a) Classical $P^{c}_{\pm 1}(b)$ and (b) quantum mechanical $P^{qm}_{\pm 1}(\theta)$ orientation parameters for $5p_{+1} \rightarrow 3p$ right-hand capture scattering, for selected impact energies $E$.}
% According to the velocity matching and propensity rules 
%models, $P^{c,qm}$ should approach +1 and -1, respectively, for all $b$'s and $\theta$'s.}
\label{fig4}
\end{figure}

Even though we do not resolve experimentally the $m$-contributions to a final $(l,m)$ capture channel, we can take adavantage of the reliability
of the MOCC calculations previously established to gauge how simple and intuitive pictures of orientation effects hold in the present low-energy
charge exchange scattering. A first (and basically classical) picture is based on a velocity-matching model which predicts that capture preferentially occurs when the
current flow of the target's valence electron is concurring with the velocity direction of the passing ion \cite{Hertel84,Campbell91}.  
Starting from Rb($5p_{+1}$), this would mean that right-hand collisions 
should be favoured and produce Na($3p_{+1}$) within the $3p$ shell -- see Fig. \ref{orientation}. An alternative picture relies on the analysis of the phases entering the
quantum mechanical scattering amplitudes and prescribes that capture preferentially occurs if stationarity of the phases is possible upon integration on $b$. 
This gave rise to the so-called propensity rules \cite{Nielsen87,Hansen92} which would give advantage to the Rb($5p_{+1}$) $\rightarrow$ Na($3p_{-1}$) capture process for right-hand collisions. 
In order to discriminate between these (contradictory) predictions, we compute the state-resolved orientation parameter
\begin{equation}
P^{qm}_{\pm 1}(\theta)=\frac{|T_{p_{+1}\rightarrow p_{+1}}(\theta,\pi)|^2-|T_{p_{+1}\rightarrow p_{-1}}(\theta,\pi)|^2}
{|T_{p_{+1}\rightarrow p_{+1}}(\theta,\pi)|^2+|T_{p_{+1}\rightarrow p_{-1}}(\theta,\pi)|^2}
\label{Pqm}
\end{equation}
which is essentially quantum mechanical in the sense that the scattering amplitudes $T$ underly integration on $b$ and therefore allow for interferences between 
classical trajectories with distinct $b$'s \cite{Hansen92}. A classical counterpart to $P^{qm}_{\pm 1}$, $P^{c}_{\pm 1}$, can be derived by 
replacing the scattering amplitudes $T$ in (\ref{Pqm}) by the transition amplitudes $<\phi_{\pm 1}|\Psi>$ issued from the MOCC calculations, where 
$\Psi$ is the total wavefunction at the end of the collision with fixed $v$ and $b$ and $|\phi_{\pm 1}>$ are the Na($3p_{\pm 1}$) oriented final states. 
$P^{c}_{\pm 1}(b)$ is computed independently for each $b$, it does not allow trajectory interplay. $P^{qm}_{\pm 1}(\theta)$ and $P^{c}_{\pm 1}(b)$ are represented
in Fig. (\ref{fig4}) for various impact energies ranging from 1 to 100 keV. For $E<20$ keV, our results show that neither the velocity matching nor the
propensity rules, which would respectively lead to $P_{\pm 1}=+1$ and -1, apply. In this regime, the impact velocity is low enough to let the electron adapt 
almost adiabatically to the nuclear motion so that the velocity matching criterion is not decisive. On the other hand, it is known that stationarity does
not drastically determine the propensity in singly-charged systems \cite{Hansen92}. However, as $E$ increases, $v$ approaches the target's valence electron velocity, 
and it is clearly seen from both Figs. \ref{fig4}(a,b) that the velocity matching prevails. This is quite satisfying as this last model draws near our
intuition of the charge exchange dynamics. Nevertheless we reiterate the conclusions of \cite{Hansen92}: care must be taken when examining orientation effects and related 
dynamics from the classical perspective at low $E$; $P^{qm}_{\pm 1}(\theta)$ and $P^{c}_{\pm 1}(b)$ cannot be linked quantitatively through the $E\theta \leftrightarrow 1/b$
correspondence previously mentioned, as revealed by the comparison of Figs \ref{fig4}(a) and (b). Nuclear trajectory interferences blur this correspondence. 

To sum up, a MOTRIMS setup have been modified in order to address the case of low-energy atomic collisions with oriented targets. Asymmetry in the DCS's, and related 
coherence properties, have then been observed with unprecedented resolution, not only for the main but also for secondary charge exchange processes. 
The measurements have revealed the importance of high-order projectile-target interactions, beyond the usual Coulombian description. This has enabled to improve the
theoretical description but also marked out the limits of the SAE and frozen core approximations. We further have shown that none of the intuitive pictures 
of electron transfer dynamics apply at low impact energies, even though the velocity matching prevails as the impact velocity approaches the target electron's one. 
Finally, we want to emphasize that the improved MOTRIMS protocol is amenable to other scattering dynamics and particle impact. We venture that such applications would enhance the experimental/theoretical synergy in many other fields than the prototypical electron capture case herein observed.
%
%Efforts must now be done in order to resolve the $m$-distributions within final $(l,m)$ multiplets, by means of a photon-particle coincidence technique. This is very 
%challenging with a confined MOTRIMS apparatus but it will open the way to separate observations of amplitudes and phases with hitherto unsurpassed accuracy. 
%Finally, the improved MOTRIMS protocol is amenable to other scattering dynamics and particle impact. We venture that this would enhance the experimental/theoretical
%synergy in many other fields than the prototypical electron capture case herein observed. 

\bibliography{biblio}

\begin{thebibliography}{999}
\bibitem{note} In practice, the left, up, right and down contributions to the DCS are 
obtained by integrating $\sigma_{p_{+1} \rightarrow 3p}(\theta,\varphi)$ for fixed $\theta$ over the respective quadrants 
$\varphi=[-\pi/4,\pi/4],\ [\pi/4,3\pi/4],\ [3\pi/4,5\pi/4]$ and $[5\pi/4,7\pi/4]$ (see Fig. \ref{orientation}). 
This considerably reduces the statistical noise in the measured profiles, but we have verified that
the integrated contributions ressemble those obtained for $\varphi=0$, $\pi/2$, $\pi$ and $3\pi/2$, respectively.
\bibitem{polarizabilities} The values $\alpha_d^{Na^+}=0.9965$ a.u. and $\alpha_q^{Na^+}=0.376$ a.u. are taken from \cite{Lombardi85} while
$\alpha_d^{Rb^+}=9.245$ a.u. and $\alpha_q^{Rb^+}=35.41$ a.u. are respectively issued from \cite{Coker76} and \cite{Johnson83}.
\end{thebibliography}

\end{document}